\documentclass[review]{elsarticle}

\usepackage{hyperref}

\journal{ArXiv}

\begin{document}

\begin{frontmatter}

\title{Implanting Synthetic Lesions for Improving Liver Lesion Segmentation in CT Exams}

\author{Dário Augusto Borges Oliveira}
\address{IBM Research, Rua Tutoia, 1157, Sao Paulo, 04007-900, Brazil}


\cortext[mycorrespondingauthor]{Corresponding author}
\ead{dariobo@br.ibm.com}


\begin{abstract}
The success of supervised lesion segmentation algorithms using Computed Tomography (CT) exams depends significantly on the quantity and variability of samples available for training. While annotating such data constitutes a challenge itself, the variability of lesions in the dataset also depends on the prevalence of different types of lesions. This phenomenon adds an inherent bias to lesion segmentation algorithms that can be diminished, among different possibilities, using aggressive data augmentation methods. In this paper, we present a method for implanting realistic lesions in CT slices to provide a rich and controllable set of training samples and ultimately improving semantic segmentation network performances for delineating lesions in CT exams. Our results show that implanting synthetic lesions not only improves (up to around 12\%) the segmentation performance considering different architectures but also that this improvement is consistent among different image synthesis networks. We conclude that increasing the variability of lesions synthetically in terms of size, density, shape, and position seems to improve the performance of segmentation models for liver lesion segmentation in CT slices.
\end{abstract}

\begin{keyword}
Lesion Synthesis \sep Liver Lesion Segmentation \sep Deep Learning
\end{keyword}

\end{frontmatter}


\section{Introduction}

Liver cancer is a very deadly type of cancer, often reported in metastasis. As in different types of cancer, early detection is crucial for treatment success, and the visual inspection of CT images is the typical detection procedure. Automated classification of liver lesions in CT exams is a challenging task, primarily due to the high variability of shape, density, location, and heterogeneity observed in such lesions. 

In this context, lesion image synthesis can be useful to provide synthesized annotated data, and different reviews for lesion synthesis in the literature have approached this topic \cite{review1,review2}. Current state-of-art for lesion synthesis relies on generative adversarial networks (GANs), as reported in \cite{medgan,review3}.

In this paper, we propose to use conditional GANs for diverse liver lesion synthesis in CT slices using controllable input data using shape, average intensity, location, and sharp edges. Our results show that different networks can generate somewhat realistic lesions from newly defined inputs. We also provide a study of the positive effect of augmenting training sets for semantic segmentation of liver lesions in CT using such synthesis schema. We report consistently improved results using two different semantic segmentation networks in the LIST challenge segmentation task. To the best of our knowledge, this is the first paper to combine controllable lesion synthesis and a stochastic framework for data augmentation through lesion implantation to improve the segmentation of real liver lesions in CT slices. 

\section{Methodology}
\label{sec:method}

Our method is composed of three steps presented in the following: creating a database of real liver lesions and the corresponding conditional images for training the synthesis networks, training conditional GANs to map between mask oriented images and real images, and implanting synthetically generated lesions in liver CT slices for improving liver lesion segmentation through stochastic data augmentation. 

\subsection{Creating Conditional Lesion Training Samples} 

To enable a stochastic yet controllable synthesis of lesions, we decided to use conditional GANs, which use paired images for translating images to a different domain by observing several pairs of translations. In our case, we aim at mapping a mask or sketch of liver lesions into lesions, and our first step is to define how this conditional image to synthesize real lesions should be. 

Some attributes were essential for allowing some control in the synthesis. The procedure should enable modifying the nodule shape, size, average intensity, location, and some structural information. That would enable controlling the synthesis and create lesion samples falling off the original dataset distribution of lesions towards the theoretical, non-achievable, theoretical distribution of lesions.   

\begin{figure}
\centering
\includegraphics[width=\textwidth]{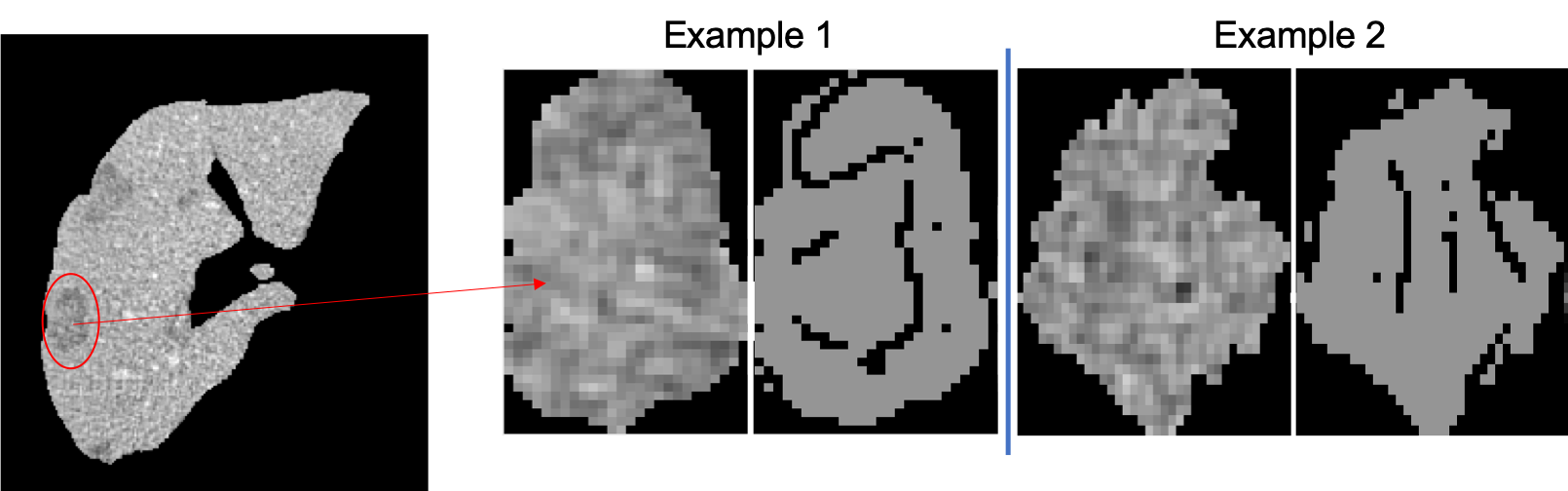}
\caption{Examples of paired samples for training nodule synthesis: target real lesion image in the left, source mask in the right. The source image holds information about shape, average value and strong edges, and the synthesis networks learn to translate source images into target real lesion images.} \label{paired-images}
\end{figure}

While a binary mask enables modifications in shape, size, and location, average intensity values, and sharp Canny edges allow learning some of the lesions' internal structure. All this information can be easily extracted from real lesion images, and building the paired dataset is straightforward with simple image processing filters, such as thresholding, averaging, and edges detection, as depicted in Figure \ref{paired-images}.

\subsection{Lesion Synthesis Using Conditional GANs} 

Having created the paired dataset, we explored two different approaches for conditional synthesis using generative adversarial networks: a mature baseline and a more recent approach.

Generative Adversarial Networks (GANs) have been widely employed in the computer vision community in the last few years. They are composed of two networks: a generator ($G$) that outputs synthesized images $y$, and a discriminator ($D$) that determines if an input image is synthesized or a real one. Such networks are trained in an adversarial scheme: while $G$ tries to learn how to produce realistic images to fool $D$, $D$ tries to discriminate between synthesized and real images correctly. Formally, given any data distribution $p_{data}(x)$, the generator $G$ learns a distribution $p_{model}(w)$ such that the discriminator can hardly distinguish between samples coming from $p_{data}(x)$ and $p_{model}(w)$. 

Conditional GANs (cGANs) are an extension of GANs where the input to the discriminator consists of samples from two domains ($x$ and $y$), and the generator synthesizes samples from one of those domains (say $y$). The loss function for conditional GANs is expressed by Equation~\ref{equ:cgansobjective}.
    
\begin{equation} \label{equ:cgansobjective}
    \mathcal{L}_{cGAN}(G,D) = E_{x,y\sim p_{data}(x,y)}[log D(x,y)]+E_{x\sim p_{(x)},z\sim p_{(z)}}[log(1 - D(x,G(x,z))]
\end{equation}
    
The solution to Equation~\ref{equ:cgansobjective} is implemented by training the generator $G$ and discriminator $D$ alternately. The discriminator is trained with images produced by the last trained generator with real images. It learns to discriminate between real and synthesized image pairs. The generator is trained using the outcome of the last trained discriminator and learns to synthesize realistic images. At the end of several training cycles, the generator is supposedly capable of producing images that the discriminator is not able to distinguish from real ones.

Pix2Pix, proposed by Isola et al \cite{pix2pix}, was used as the baseline, and SPADE, that report state-of-art results for synthesis from labeled masks \cite{spade}, was used for comparison. Both architectures were trained using the same paired datasets, as detailed in the experiments section.

\subsection{Implanting Synthetic Lesions for Data Augmentation}

As exposed in the introduction, our primary assumption in this study is that synthetically increasing the variability of liver lesions can help to boost the performance of semantic segmentation networks for lesion delineation. The strategy proposed for assessing this assumption is to implant synthetically generated lesions in existing CT liver slices and map them into the respective masks. With a trained network for lesion synthesis, we can vary the location, size, shape, average value, and internal edges of masks to create a rich dataset of realistic synthetic lesions for different slices. 

\begin{figure}
\includegraphics[width=\textwidth]{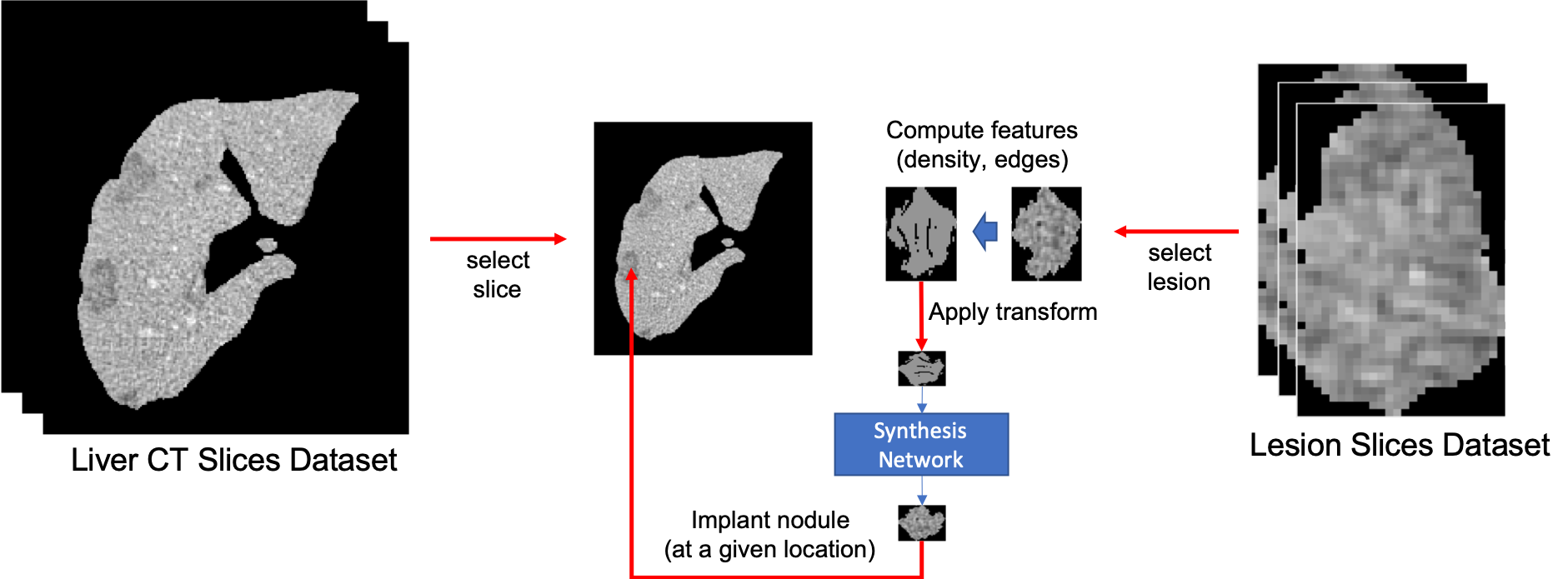}
\caption{The lesion implant strategy consists of selecting a liver slice and implanting synthesized lesions iteratively. Each lesion is transformed at random to increase the variability of lesion patterns, and the location to be implanted is also chosen at random.} \label{nodule-implant}
\end{figure}

\begin{figure}
\centering
\includegraphics[width=0.7\textwidth]{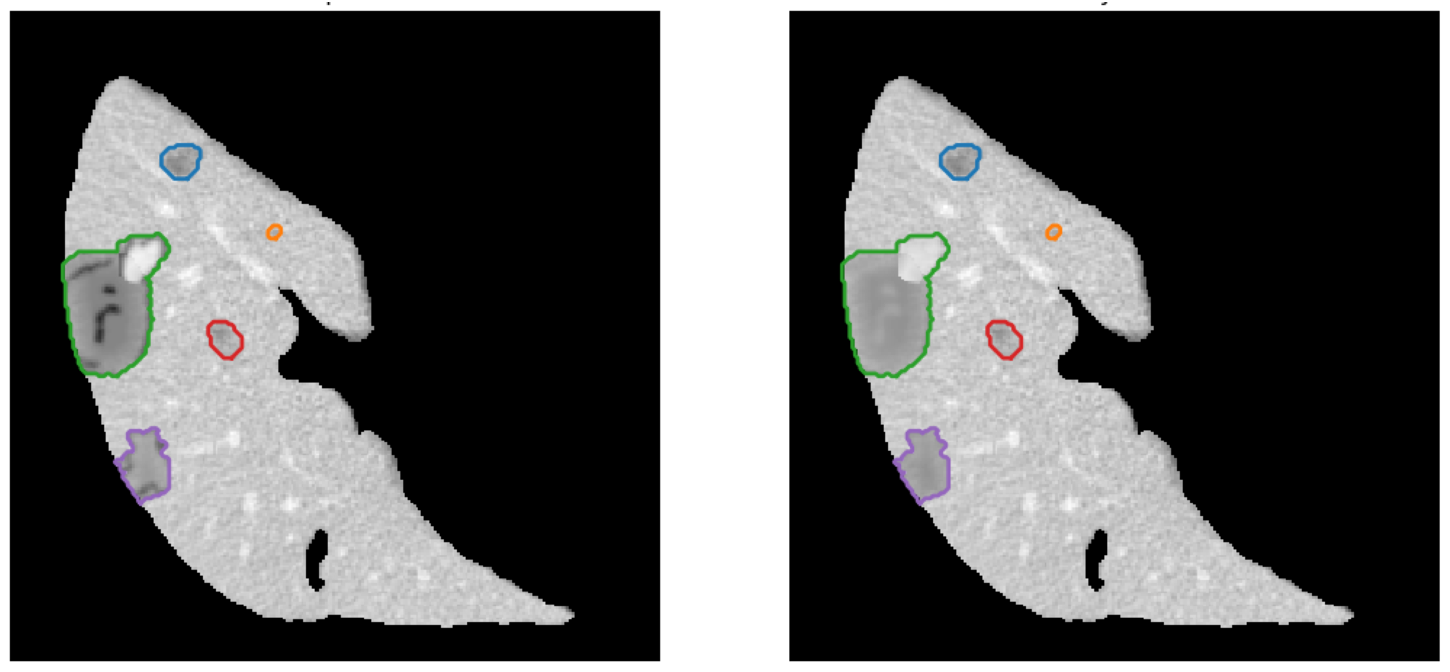}
\caption{One can notice different lesions, synthetic and real, with different coloured borders: green and purple borders indicate synthetic lesions, and the other coloured borders real ones. In the left the liver CT slice with masks created at random, in the right the resulting slice with synthesized nodules.} \label{implant-examples}
\end{figure}

Figure \ref{nodule-implant} illustrates the process. We select a liver CT slice and several lesions for implanting. For each lesion selected at random from the training set, we apply random rotation, translation, resizing, average value, and create a realistic synthetic lesion corresponding to this input modified mask. An example of an outcome for some lesions can be observed in Figure \ref{implant-examples}. 

\section{Experiments}

In our experiments, we target at reporting the impact of our data augmentation approach in different semantic segmentation networks. First, we evaluate the quality of lesion synthesis, and then we evaluate the impact of enriching the variability of the lesions training set in semantic segmentation networks using our approach.

\subsection{Lesion Synthesis} 

To train networks for lesion synthesis, we used the public challenge LITS dataset \cite{lits}, which sets 100 exams for training and validation, and a testing set with 30 exams used with no modification for evaluating the semantic segmentation network performances. The training set comprises a total of around 10.000 liver CT slices and 4.400 samples of lesion CT slices. 

Each lesion CT slice in training set derived the corresponding mask using the transformations depicted in section \ref{sec:method} and the same data used to train both Pix2Pix and SPADE networks. 

For training Pix2Pix, we run 150 epochs using Adam optimizer with an initial learning rate of 0.0002 and a momentum of 0.5. The parameters for the generator loss function were set to GAN weight=1 and L1 weight=100. For training SPADE, we run 150 epochs using Adam optimizer with an initial learning rate of 0.0002 and a momentum of 0.9. The parameters for loss functions were set as the default proposed in the paper.

\subsection{Lesion Segmentation}

Each trained network was used for synthetically implanting lesions in existing liver CT slices, up to the same amount available in original training data. That allows us to compare the performance of models trained using only data synthesized using Pix2Pix or SPADE with the performance achieved using only real data, or with the combination of real and synthetic data (Pix2Pix+ and SPADE+). 

The lesion segmentation experiments involved two different semantic segmentation networks: U-Net \cite{unet} as a baseline, and PSP-Net \cite{pspnet} that reports state-of-art results for semantic segmentation \cite{review4}. Our experiments considered the simplest and toughest setup possible: input images were bi-dimensional CT slices with lesions, and target was the corresponding lesion masks. Since we were not using 3D or 2.5D networks for segmentation, neighboring information was not used for easing the segmentation process whatsoever. 

Five different experiments were performed for each semantic segmentation network depending on the training set used, and results are discussed in the following section: 
\begin{enumerate}
    \item Only Pix2Pix-generated lesions.
    \item Only SPADE-generated lesions.
    \item Only real lesions.
    \item A combination of Pix2Pix-generated and real lesions.
    \item A combination of SPADE-generated and real lesions.
\end{enumerate}

\section{Results and Discussion}

We report our results from two perspectives: the quality of lesion synthesis and the impact of synthetically implanting lesions in liver CT slices to improve the performance of segmentation networks. 

\subsection{Lesion Synthesis} 

\begin{figure}
\centering
\includegraphics[width=1\textwidth]{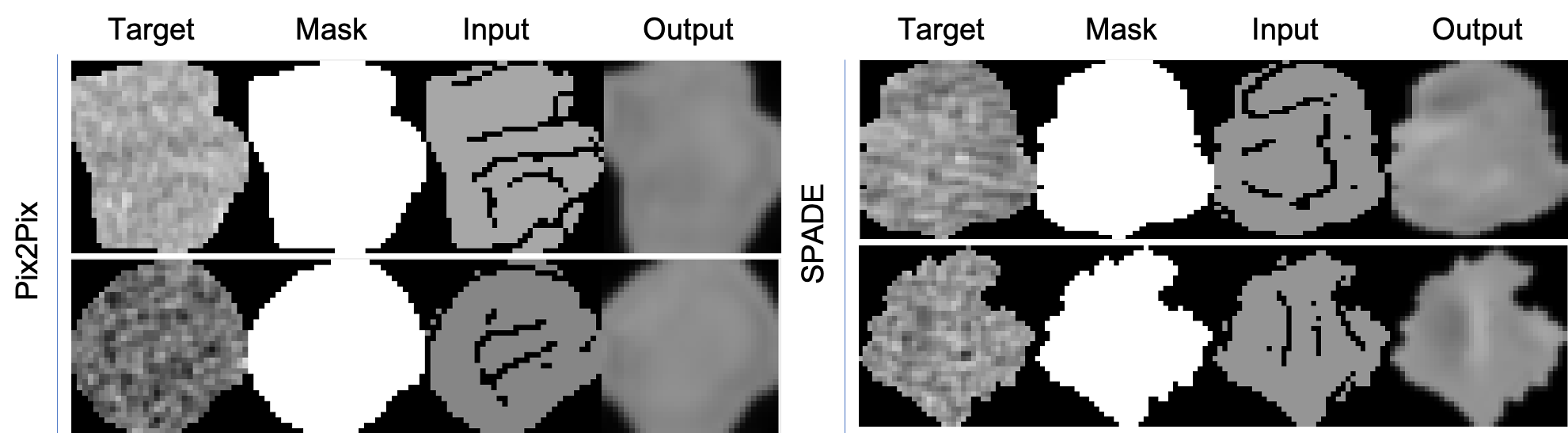}
\caption{Lesion synthesis examples for Pix2Pix and SPADE cGANs. One can notice that SPADE produces visually more appealing results, that follows an internal structure defined by edges in the input data. Still, the outcome struggles to accurately reproduce the texture observed in target real lesions.} \label{synthesis-examples}
\end{figure}

Considering the lesion synthesis, even if we observed consistently more visually appealing results in SPADE results in comparison to Pix2Pix baseline - as shown in Figure \ref{synthesis-examples}, the mean squared error in comparison to the real lesion was virtually the same in both networks: $0.0108$ for Pix2Pix and $0.0102$ for SPADE, slightly better.

One can observe that SPADE can reconstruct more information from the dull, sharp edges provided in the source images. It is noticeable, however, that both networks struggle to reconstruct texture patterns for the lesion with the provided source information. 

\subsection{Lesion Segmentation}

The impact of a more diverse training set to the segmentation networks seems consistent and encouraging. We observed in both U-Net and PSP-Net a gain in performance with the use o synthetic nodules implantation in liver CT slices: up to \textbf{6\%} difference in U-Net and \textbf{2\%} in PSP-Net, which represents a gain of around \textbf{12\%} and \textbf{3.5\%} respectively.   

More interestingly, we observed the same increase in performance using only synthetic data for both synthesis networks, or a combination of real and synthetic lesions. We believe the lesion synthesis was good enough for challenging the training, but most importantly that the stochastic process of implanting lesions at a different location, size, and average intensity helped the network to increase its robustness against variability in these features.


\begin{table}
\centering
\caption{Liver lesion segmentation Dice score for different training sets, using only synthetic data for Pix2Pix and SPADE, using only original data, using a combination of Pix2Pix or SPADE and original data}\label{tab1}
\begin{tabular}{|c|c|c|c|c|c|}
\hline
 &  { Pix2Pix }  &  { SPADE }  &  {   Original   }  & { Pix2Pix+ } &  SPADE+ \\
\hline
{U-Net} & \textbf{0.568} & 0.5604 & 0.5038 & 0.565 & 0.5607  \\
\hline
{PSP-Net} & \textbf{0.6089} & 0.6055 & 0.5843 & 0.6082 & 0.605 \\
\hline
\end{tabular}
\label{table-results}
\end{table}

While the synthesis network choice does not seem to have an impact on the segmentation performance, both segmentation networks benefit from having a more diverse training set for delineating lesions in the testing set, as depicted in table \ref{table-results}. We believe this positive impact is mainly related to the variability in terms of size, shape, and position of nodules implanted for training the segmentation network.

\section{Conclusions}

This paper presents a methodology for improving segmentation network performances for liver lesions by synthetically implanting lesions in CT liver slices. Controlling parameters such as lesion size, shape, location, internal structure, and average intensity, we were able to enrich the training set of semantic segmentation networks and boost their performance in up to 12\%. First, we create a paired dataset of liver lesions that allowed the training of image translation networks that map modified masks to real lesions. Then, using the trained network for lesion synthesis, we created a dataset for training semantic segmentation networks, by stochastic implanting lesions in liver CT slices. 

Our results show the potential of such a methodology for improving liver lesion segmentation. We report consistent improvements considering different network synthesis and for different semantic segmentation networks, only by increasing the variability of liver lesion training set synthetically. In the best configuration tested, we achieved an increase in performance of around 12\% of dice-score. The gains were also consistent: all configurations achieved slightly better results in comparison to the network trained using only original data.

As further research, we envisage using this synthesis methodology for implanting nodules in 3D, evaluating the separation of shape and style in the synthesis process to control them separately within the network and testing other synthesis networks for more realistic nodules textures. 

\bibliography{elsarticle-template}

\end{document}